\documentclass[10pt,conference]{IEEEtran}

\usepackage{flushend}
\usepackage{lipsum}
\usepackage{stix}
\usepackage{cite}
\usepackage{amsmath,amssymb,amsfonts}
\usepackage{algorithmic}
\usepackage{graphicx}
\usepackage{textcomp}
\usepackage{xcolor}
\def\BibTeX{{\rm B\kern-.05em{\sc i\kern-.025em b}\kern-.08em
    T\kern-.1667em\lower.7ex\hbox{E}\kern-.125emX}}

\usepackage{url}
\usepackage{hyperref}
\hypersetup{
    colorlinks=true,
    citecolor=black,linkcolor=black,
    filecolor=black, urlcolor=cyan,
}

\newcommand{\eg}{\emph{e.g.,}}
\newcommand{\ie}{\emph{i.e.,}}
\newcommand{\etal}{\emph{et al.}}

\usepackage{tcolorbox}
\usepackage{dblfloatfix}
\usepackage{booktabs}

\newtcolorbox{boxH}{
    colback = white!90!gray, 
    colframe = black, 
    boxrule = 0pt, 
    leftrule = 3pt
}

\usepackage{listings}
\definecolor{codegreen}{rgb}{0,0.6,0}
\definecolor{codegray}{rgb}{0.5,0.5,0.5}
\definecolor{codepurple}{rgb}{0.58,0,0.82}
\definecolor{backcolour}{rgb}{0.95,0.95,0.92}
\lstdefinestyle{mystyle}{
    commentstyle=\color{codegreen},
    keywordstyle=\color{magenta},
    numberstyle=\tiny\color{codegray},
    stringstyle=\color{codepurple},
    basicstyle=\ttfamily\footnotesize,
    breakatwhitespace=false,         
    breaklines=true,                 
    captionpos=b,                    
    keepspaces=true,                 
    numbers=left,                    
    numbersep=5pt,                  
    showspaces=false,                
    showstringspaces=false,
    showtabs=false,                  
    tabsize=2
}

\begin{document}

\title{What Do Contribution Guidelines Say About Software Testing?}


 \author{
 \IEEEauthorblockN{Bruna Falcucci, Felipe Gomide, Andre Hora}
 \IEEEauthorblockA{
\textit{Department of Computer Science, UFMG}\\
 Belo Horizonte, Brazil \\
 {\{brunafalcucci, felipe.gomide, andrehora\}}@dcc.ufmg.br}
 }

\maketitle

\begin{abstract}
Software testing plays a crucial role in the contribution process of open-source projects.
For example, contributions introducing new features are expected to include tests, and contributions with tests are more likely to be accepted.
Although most real-world projects require contributors to write tests, the specific testing practices communicated to contributors remain unclear.
In this paper, we present an empirical study to understand better how software testing is approached in contribution guidelines.
We analyze the guidelines of 200 Python and JavaScript open-source software projects.
We find that 78\% of the projects include some form of test documentation for contributors.
Test documentation is located in multiple sources, including \texttt{CONTRIBUTING} files (58\%), external documentation (24\%), and \texttt{README} files (8\%).
Furthermore, test documentation commonly explains how to run tests (83.5\%), but less often provides guidance on how to write tests (37\%).
It frequently covers unit tests (71\%), but rarely addresses integration (20.5\%) and end-to-end tests (15.5\%).
Other key testing aspects are also less frequently discussed: test coverage (25.5\%) and mocking (9.5\%).
We conclude by discussing implications and future research.
\end{abstract}



\section{Introduction}

Open-source software projects typically provide \emph{contribution guidelines} detailing how to submit contributions~\cite{contributor_guidelines, elazhary2019not, open_source_guides}.
For maintainers, contribution guidelines communicate how developers should contribute.
For contributors, these guidelines support the verification of well-formed and valuable contributions.
Both project maintainers and contributors benefit from saving time and avoiding problems caused by inadequately submitted contributions~\cite{contributor_guidelines}.
This kind of documentation may also alleviate barriers faced by new contributors~\cite{steinmacher2015systematic}.
Due to its importance, GitHub recommends having contribution guidelines as a prerequisite for launching open source projects~\cite{open_source_guides} and provides guides to support their creation~\cite{contributor_guidelines, elazhary2019not}.

Contribution guidelines may have multiple contents detailing the workflow,  acceptance criteria, or how to create forks, branches, pull requests, and more~\cite{elazhary2019not}.
Due to its importance to software development, software testing is also a fundamental part of the contributing process~\cite{elazhary2019not, tsay2014influence, pham2013creating, dabbish2012social}.
Indeed, 52\% of the contribution guidelines state that contributions have to include test cases~\cite{elazhary2019not}.
Contributions introducing new features are expected to include tests~\cite{pham2013creating}, and core maintainers look for the inclusion of test cases as a signal of the thoroughness of the contribution~\cite{dabbish2012social}.
Consequently, contributions that include test cases are more likely to be accepted~\cite{tsay2014influence}.


GitHub recommends that contribution guidelines might require tests to improve the quality: ``\emph{Tests help contributors feel confident that they won’t break anything. They also make it easier for you to review and accept contributions quickly}''~\cite{open_source_guides}.
Similarly, popular guides with best practices recommend that contributions should at least detail how to run the tests~\cite{makeareadme}.
It is not surprising that open-source projects treat tests seriously in their contribution guidelines.
For example, the Pandas contribution documentation states \emph{``Ensure you have appropriate tests. These should be the first part of any PR''},\footnote{\url{https://pandas.pydata.org/docs/development/contributing.html}} while the Flask one notes: \emph{``Include tests if your patch adds or changes code''}.\footnote{\url{https://github.com/pallets/flask/blob/main/CONTRIBUTING.rst}}
Other projects go a step further by providing detailed information about testing in contribution guidelines.
For instance, Cypress provides an overview of the test types contributors may create, such as unit, integration, and e2e.\footnote{\url{https://github.com/cypress-io/cypress/blob/develop/CONTRIBUTING.md}}
The CPython project has extensive contribution documentation on how to run/write the tests and how to increase test coverage.\footnote{\url{https://devguide.python.org/testing/run-write-tests}}


Despite software testing's critical role in contribution guidelines, we are not yet aware of how software testing is explained to contributors.
While most real-world projects require developers to write tests in contributions~\cite{elazhary2019not}, it is unclear whether these projects actually communicate testing practices to support contributions.
Furthermore, in the case projects do communicate, it is unclear what testing practices are presented for contributors.
For example, maintainers may focus on various topics, including run/write tests, test types, coverage, mocking, and other software testing practices.
Assessing this information can provide the basis for unveiling communication patterns (that should be adopted in contribution guidelines) and anti-patterns (that should be avoided).

In this paper, we present an empirical study to understand better how software testing is approached in the contribution guidelines of open-source software projects.
We analyze the contribution guidelines of 200 popular open-source projects written in Python and JavaScript.
We propose to answer the following research questions:

\begin{itemize}

    \item \textbf{RQ1: How frequently is test documentation for contributors?}
    We find that 78\% of the projects include some form of test documentation for contributors.
    Test documentation is mostly located in \texttt{CONTRIBUTING} files (58\%) and external documentation (24\%), but also occurs in \texttt{README} files (8\%).

    \item \textbf{RQ2: What is the content of test documentation for contributors?}
    Test documentation commonly explains how to run tests (83.5\%), but less often provides guidance on how to write tests (37\%).
    Moreover, it frequently covers unit tests (71\%), but rarely addresses integration (20.5\%) and end-to-end tests (15.5\%).
    Other key testing aspects are also less frequently discussed: test coverage (25.5\%), mocking (9.5\%), and best practices (9\%). 
    
\end{itemize}

\noindent\emph{Contributions.}
The contributions of this paper are twofold.
First, we provide an initial empirical study to assess test documentation for contributors.
Second, we discuss practical implications for researchers and practitioners.

\section{Study Design}
\label{sec:design}

\subsection{Case Study}

We aim to study the contribution guidelines of real-world and relevant software projects.
For a better perspective of the contribution guidelines landscape, we focus on projects written in the two most popular programming languages nowadays: Python and JavaScript.
For each programming language, we relied on the GitHub Search tool (GHS)~\cite{msr2021data} to collect the top-100 most popular software projects hosted on GitHub according to the number of stars (a metric primarily adopted in the software mining literature as a proxy of popularity~\cite{hudson16, jss-2018-github-stars}).
In this process, we took special care to filter out non-software projects, such as tutorials, examples, and code samples.
Our final dataset is composed of 200 software projects (100 in Python and 100 in JavaScript).
On the median, the selected projects have 35.7K stars.

\subsection{Detection of Contribution Guidelines}

GitHub recommends to write contribution guidelines in \texttt{CONTRIBUTING} files~\cite{contributor_guidelines}, as in jQuery.\footnote{\url{https://github.com/jquery/jquery/blob/main/CONTRIBUTING.md}}
Contribution guidelines may also be written directly in \texttt{README} files~\cite{prana2019categorizing}, as in Webpack.\footnote{\url{https://github.com/webpack/webpack/blob/main/test/README.md}}
More comprehensive guidelines may be located in external documentation, as in Black\footnote{\url{https://black.readthedocs.io/en/latest/contributing}} and CPython.\footnote{\url{https://devguide.python.org}}


In this study, we consider all kinds of contribution guidelines as long as they are dedicated to contributors.
We look for contribution guidelines (or links) in \texttt{README} and \texttt{CONTRIBUTING} files as well as other variations, such as \texttt{DEVELOPMENT} files.
We verify not only the root directory of the projects but also the source code and test folders.
Following these steps, we found that most projects have contribution guidelines (184 out of 200).
This is expected, considering the relevance and popularity of the selected projects.

All selected projects, contribution guidelines, and extracted information are publicly available in our dataset at \url{https://doi.org/10.5281/zenodo.14046800}.

\subsection{Detection of Test Documentation for Contributors}

Next, we searched the contribution guidelines for any test documentation.
We define \emph{test documentation for contributors} as any information related to software testing that is aimed at contributors and is included in contribution guidelines.
Particularly, we focused on essential testing topics such as how to run or write tests, best practices, test coverage, mocking, and test types (\ie~unit, integration, and end-to-end)~\cite{open_source_guides, makeareadme}.
If no test documentation is found for a given project, a second author conducts a follow-up search to double-check and minimize the risk of false negatives.
Following these steps, we detected that 156 out of 200 projects include test documentation for contributors.



\subsection{Research Questions}

We propose two RQs to assess the test documentation for contributors.
First, we explore the \emph{frequency} of test documentation for contributors.
The rationale is to better understand the extent of the phenomenon under study and to determine whether any differences exist between Python and JavaScript.
Second, we explore the \emph{content} of test documentation for contributors.
The rationale is to uncover which testing topics are addressed in contribution guidelines and determine which are the most common and lacking.
Such information may help discover best practices and guidelines for creating test documentation for contributors.

\section{Results}

\subsection{RQ1: Frequency of test documentation for contributors}

We find test documentation for contributors in 78\% of the analyzed projects (156 out of 200), as detailed in Table~\ref{tab:summary}.
Notice that Python has slightly more systems with test documentation (82 in 100) than JavaScript (74 in 100).

\begin{table}[h]
    \centering
    \footnotesize
    \caption{Summary of test documentation for contributors.}
    \begin{tabular}{l rrr}
        \toprule
         & \textbf{Python} & \textbf{JavaScript} & \textbf{Total} \\ \midrule
        Analyzed Projects  & 100 & 100 & 200 \\ \midrule
        $\smblkcircle$ With Test Documentation (\#)  & 82 & 74 & 156 \\ 
        $\smblkcircle$ With Test Documentation (\%)  & 82\% & 74\% & 78\% \\
        \bottomrule
    \end{tabular}
    \label{tab:summary}
\end{table}

Table~\ref{tab:location} presents the location of the 156 test documentation.
We see that most test documentation for contributors is naturally located in the recommended \texttt{CONTRIBUTING} files (58\%).
Test documentation is also present in external documentation (24\%) and \texttt{README} files (8\%).
We also find test documentation in other files (10\%), such as \emph{developers}, \emph{development}, \emph{testing}, and \emph{workflow}.
Overall, there is no major difference between Python and JavaScript regarding the location.
The sole discrepancy happens in the location external documentation: Python has 26 in this location, while JavaScript has only 12.
This may be explained by the fact that Python has the \emph{Read the Docs}\footnote{\url{https://about.readthedocs.com}} platform to build documentation (6 out of the 26 Python systems rely on it).

\begin{table}[h]
    \centering
    \footnotesize
    \caption{Location of test documentation for contributors.}
    \begin{tabular}{l rrrr}
        \toprule
        \textbf{Location} & \textbf{Python} & \textbf{JavaScript} & \textbf{Total} & \textbf{\%} \\ \midrule
        \texttt{CONTRIBUTING} file  & 41 & 49 & 90 & 58\% \\
        External documentation      & 26 & 12 & 38 & 24\% \\
        \texttt{README} file        & 5 & 7 & 12 & 8\% \\
        Other files                 & 10 & 6 & 16 & 10\% \\ \midrule
        All                         & 82 & 74 & 156 & 100\% \\
        \bottomrule
    \end{tabular}
    \label{tab:location}
\end{table}

\begin{boxH}
\textbf{Finding 1}:
\textbf{(i)} 78\% of the projects include some form of test documentation for contributors.
\textbf{(ii)} Test documentation is mostly located in \texttt{CONTRIBUTING} files (58\%) and external documentation (24\%), but also occurs in \texttt{README} files (8\%).
\end{boxH}


\subsection{RQ2: Content of test documentation for contributors}

Table~\ref{tab:content} summarizes the content of the test documentation for contributors. 
It is divided into five categories: (1) how to run/write tests, (2) test types, (3) test coverage, (4) mocking, and (5) best practices \& tips.
The most frequent content is \emph{how to run/write tests} (present in 86\% of the test documentation), followed by  \emph{test types} (74\%) and \emph{test coverage} (25.5\%).
Next, we briefly describe each category.

\begin{table}[h]
    \centering
    \footnotesize
    \caption{Content of test documentation for contributors.}
    \begin{tabular}{l rrrr}
        \toprule
        \textbf{Content} & \textbf{Python} & \textbf{JavaScript} & \textbf{Total} & \textbf{\%}  \\ \midrule
        \textbf{How to run/write tests} & 72 & 62 & 134 & 86\% \\
        $\smblkcircle$ How to run tests & 70 & 60 & 130 & 83.5\% \\
        $\smblkcircle$ How to write tests & 30 & 28 & 58 & 37\% \\ \midrule
        \textbf{Test types} & 65 & 50 & 115 & 74\% \\
        $\smblkcircle$ Unit test & 64 & 47 & 111 & 71\% \\
        $\smblkcircle$ Integration test & 20 & 12 & 32 & 20.5\% \\
        $\smblkcircle$ e2e test & 7 & 17 & 24 & 15.5\% \\ \midrule
        \textbf{Test coverage} & 31 & 9 & 40 & 25.5\% \\ \midrule
        \textbf{Mocking} & 10 & 5 & 15 & 9.5\% \\ \midrule
        \textbf{Best practices \& tips} & 8 & 6 & 14 & 9\% \\
        \bottomrule
    \end{tabular}
    \label{tab:content}
\end{table}

\textbf{How to run/write tests:}
Most real-world projects require contributors to test their patches locally and write new tests if code is added or changed~\cite{elazhary2019not, open_source_guides, pham2013creating}.
Therefore, ideally, test documentation should describe at least how to run and write tests.
Overall, we find that 86\% (134 in 156) of the test documentation includes information about running and/or writing tests.
While 83.5\% describe how to run tests, only 37\% actually detail how to write tests.
The typical how-to-run test simply includes running commands, as in the test documentation of Meteor.\footnote{\url{https://github.com/meteor/meteor/blob/devel/DEVELOPMENT.md\#running-tests-on-meteor-core}}
Some test documentations go a step further, explaining how to write tests as in Next.js\footnote{\url{https://github.com/vercel/next.js/blob/canary/contributing/core/testing.md\#writing-tests-for-nextjs}} and CPython.\footnote{\url{https://devguide.python.org/testing/run-write-tests/\#writing}}
In this case, the recommendations are diverse, including how to write effective tests, deciding where to place the tests, choosing the appropriate types of tests to write, selecting a suitable testing library, and more

\textbf{Test types:}
We analyze the three test types of the test pyramid: unit, integration, and end-to-end (e2e).
Overall, 74\% (115 in 156) of the test documentation mentions at least one test type.
The most common is unit test, present in 71\% of the test documentation.
For example, the test documentation of Langchain mentions: \emph{``Unit tests: run on every pull request, so they should be fast and reliable [...] Unit tests cover modular logic that does not require calls to outside APIs. If you add new logic, please add a unit test''}.\footnote{\url{https://python.langchain.com/docs/contributing/testing}}
Integration test is found in 20.5\% of the cases.
Project Langchain also mentions: \emph{``Integration tests cover logic that requires making calls to outside APIs [...] If you add support for a new external API, please add a new integration test.''}.
Lastly, we have e2e test, which is found in 15.5\%.
For example, the test documentation of Next.js states: \emph{``End-to-end (e2e) tests are run in complete isolation from the repository''}.\footnote{\url{https://github.com/vercel/next.js/blob/canary/contributing/core/testing.md}}


\textbf{Test coverage:}
Test coverage measures the percentage of code that is covered (and uncovered) by tests~\cite{hora2023excluding} and is typically used to gauge the test effectiveness~\cite{coverage_py}.
We find that 25.5\% (40 in 156) of the test documentation mentions test coverage.
For example, CPython has specific documentation dedicated to test coverage: \emph{``Increase test coverage [...] Ideally we would like to have 100\% coverage, but any increase is a good one''}.\footnote{\url{https://devguide.python.org/testing/coverage}}
Likewise, scikit-learn also mentions a coverage threshold: \emph{``We expect code coverage of new features to be at least around 90\%''}.\footnote{\url{https://scikit-learn.org/stable/developers/index.html}}
Project Materialize details what to do with code that is hard to cover: \emph{``Try and cover as many cases as you can, but don’t worry if there are some edge cases. You can add comments describing some problematic edge cases in TODOs so we know about them.''}.\footnote{\url{https://github.com/Dogfalo/materialize/blob/v1-dev/CONTRIBUTING.md}}

\textbf{Mocking:}
When creating tests, developers may find dependencies that make the test harder to implement. In this scenario, they can use mocks (test doubles) to emulate the dependencies' behavior, contributing to making the test fast, isolated, and deterministic~\cite{meszaros2007xunit, pereira2020assessing, spadini2017mock, spadini2019mock, fazzini2022use}.
We detect that only 9.5\% of the test documentation refers to test doubles.
For example, project Vuex presents how to abstract API calls: \emph{``When testing actions, we usually need to do some level of mocking - for example, we can abstract the API calls into a service and mock that service inside our tests''}.\footnote{\url{https://vuex.vuejs.org/guide/testing.html}}
Project Pandas details what a unit test should not access: \emph{``A unit test should not access a public data set over the internet due to flakiness of network connections [...] To mock this interaction, use the httpserver fixture''}.\footnote{\url{https://pandas.pydata.org/docs/dev/development/contributing_codebase.html}}
Likewise, the OpenHands test documentation details: \emph{``When you launch an integration test, mock responses are loaded and used to replace a real LLM's response, so that we get deterministic and consistent behavior, and most importantly, without spending real money''}.\footnote{\url{https://github.com/All-Hands-AI/OpenHands/blob/main/tests/integration/README.md}}


\textbf{Best practices \& tips:}
Lastly, we looked for best practices and tips in software testing.
We find that only 9\% (14 in 156) of the test documentation provides this information.
For example, project Next.js recommends creating bug-reproducing tests: \emph{``Best Practices: When applying a fix, ensure the test fails without the fix. This makes sure the test will properly catch regressions''}.\footnote{\url{https://github.com/vercel/next.js/blob/canary/contributing/core/testing.md\#best-practices}}
The test documentation of jQuery provides tips to improve test performance: \emph{``Test Suite Tips: During the process of writing your patch, you will run the test suite MANY times. You can speed up the process by narrowing the running test suite down to the module you are testing [...]''}.\footnote{\url{https://github.com/jquery/jquery/blob/main/CONTRIBUTING.md\#test-suite-tips}}

\begin{boxH}
\textbf{Finding 2}:
\textbf{(i)} Test documentation commonly explains how to run tests (83.5\%), but less often provides guidance on how to write tests (37\%).
\textbf{(ii)} Test documentation frequently covers unit tests (71\%), but rarely addresses integration (20.5\%) and end-to-end tests (15.5\%).
\textbf{(iii)} Other key testing aspects are also less frequently discussed: test coverage (25.5\%), mocking (9.5\%), and best practices (9\%). 
\end{boxH}

\section{Discussion}


\noindent\textbf{Current status of test documentation for contributors.}
We find that test documentation for contributors has diverse but unbalanced test content (\eg~more how-to-run than how-to-write, more unit than e2e tests).
Moreover, we detect that some content, such as e2e tests, test coverage, mocking, and best practices, is underrepresented in the test documentation.
Without clear guidance, contributors must make their own decisions on critical aspects, such as when to create e2e tests, which mocking framework to use, what code to mock, and what coverage threshold to adopt.
We recommend that project maintainers include this test information to inform contributors better and prevent issues caused by inadequately submitted contributions~\cite{contributor_guidelines}.
Also, researchers can propose novel tools to automatically detect contribution guidelines that lack certain test-related content (\eg~mocking) to warn maintainers.

\smallskip



\noindent\textbf{Guidelines to create test documentation for contributors.}
Our goal in this research is to provide guidelines to help create better test documentation for contributors.
RQ2 provides an initial step in this direction.
For example, for test coverage, we found many concrete recommendations for contributors about coverage increments, coverage thresholds, code that is hard to cover, and more.
In the mock context, we found recommendations of how to mock, what to mock, and what tool to use.
Regarding the test types, we found definitions of what characterizes a specific type of test, when a test should be created, and when a test should be run.
Therefore, we envision that a deep qualitative exploration of the actual test documentation for contributors can reveal best practices and guidelines that can be reused by open-source projects.

\section{Limitations}

This study provides an initial assessment of test documentation for contributors.
The test documentation was detected manually, with the support of two authors of the paper.
As detailed in Section~\ref{sec:design}, when no test documentation was found for a given project, a second author conducted a follow-up search to double-check and minimize the risk of false negatives.
Moreover, RQ2 explored the content of such documentation.
However, we did not assess \emph{what} recommendations are actually provided for contributors.
We plan to deeply explore the recommendations in further studies.

\section{Related Work}

Multiple studies focus on better understanding the content of readme files and contribution guidelines~\cite{liu2022readme, wang2023study, prana2019categorizing, elazhary2019not, steinmacher2015systematic}.
Liu~\etal~assessed the structure of readme files in Java projects~\cite{liu2022readme}.
The authors found that the majority of readme files do not align with the GitHub guidelines, but repositories whose readme files follow the GitHub guidelines tend to be more popular.
Wang~\etal~also studied the correlation between readme files and project popularity~\cite{wang2023study}.
Prana~\etal~found that information about the ``What'' and ``How'' of a repository is common in readme files~\cite{prana2019categorizing}.
Elazhary~\etal~explored the content of contribution guidelines and compared their actual practices with the prescribed contribution
guidelines~\cite{elazhary2019not}.
The authors found that most projects diverge significantly from the expected process.
Other research shows that such documentation may support new contributors~\cite{steinmacher2015systematic}.
Despite the various studies on contribution guidelines, there is a lack of research addressing software testing in such guidelines.
Our research aims to fill this gap in the literature.

\section{Conclusion and Further Studies}

We presented an empirical study to understand better how software testing is approached in contribution guidelines of 200 Python and JavaScript projects.
We revealed the following insights:
(1) 78\% of the projects include some form of test documentation for contributors;
(2) they focus more on how-to-run rather than how-to-write;
(3) unit tests are commonly covered, while end-to-end tests receive less attention; and
(4) discussions on test coverage and mocking are rarer.

In future work, we plan to \emph{qualitatively} explore test documentation to catalog best practices and guidelines for contributors.
In addition to the studied test content, we plan to address other aspects relevant to testing that may appear in contributor guidelines, such as testing techniques.

\section*{Acknowledgments}

This research is supported by CAPES, CNPq, and FAPEMIG.

\bibliographystyle{IEEEtran}
\bibliography{main}

\end{document}